\documentclass[aps, reprint,showpacs,showkeys,pra,longbibliography]{revtex4-1}
\usepackage{dcolumn}
\usepackage[utf8]{inputenc}
\usepackage{amsmath}
\usepackage{amsfonts}
\usepackage{amssymb}
\usepackage{graphicx}
\usepackage{hyperref}
\usepackage{xcolor}

\newcommand{\rme}{\mathrm{e}}

\begin{document}

\title{Describing squeezed-light experiments without squeezed-light states}

\author{Tam\'iris R. Calixto}
\affiliation{Departamento de F\'isica, Universidade Federal de Minas Gerais, Belo Horizonte, MG 31270-901, Brazil}
\author{Pablo L. Saldanha}\email{saldanha@fisica.ufmg.br}
\affiliation{Departamento de F\'isica, Universidade Federal de Minas Gerais, Belo Horizonte, MG 31270-901, Brazil}

\date{\today}

\begin{abstract}
Coherent states are normally used to describe the state of a laser field in experiments that generate and detect squeezed states of light. Nevertheless, since the laser field absolute phase is unknown, its quantum state can be described by a statistical mixture of coherent states with random phases, which is equivalent to a statistical mixture of Fock states. Here we describe single-mode squeezed vacuum experiments using this mixed quantum state for the laser field. Representing the laser state in the Fock basis, we predict the usual experimental results without using the squeezing concept in the analysis and concluding that no squeezed state is generated in the experiments. We provide a general physical explanation for the noise reduction in the experiments in terms of a better definition of the relative phase between the signal and local oscillator fields.  This explanation is valid in any description of the laser field (in terms of coherent or Fock states), thus providing a deeper understanding of the phenomenon. 
\end{abstract}


\maketitle

\par A squeezed state of light is a quantum state in which the variance in some quadrature is smaller than the vacuum noise \cite{livroqo,livrobase,expestadocomprimido}, having many applications in quantum metrology \cite{caves81,artigoPhysReports,artigonaturepht,artigoPRLsqueezed,tse19,acernese19,zhao20,mcculler20} and quantum information \cite{teleportation,Coelho823,chen14,PhysRevApplied.3.044005,PhysRevA.98.023823,Israel:19}. The use of squeezed states allows one to achieve sub shot-noise accuracy, since the quantum noise is reduced in some quadrature, enhancing the system sensitivity \cite{livroqo,livrobase,artigoPhysReports}. One example is the sensitivity enhancement of gravitational wave detectors \cite{artigonaturepht}, and there is an intense active research for improving this sensitivity \cite{artigoPhysReports,tse19,acernese19,zhao20,mcculler20}. Entanglement is present in squeezed states of two or more modes, which permits the execution of many quantum information protocols \cite{teleportation,Coelho823,chen14,PhysRevApplied.3.044005,PhysRevA.98.023823,Israel:19}.

Squeezed states of light are usually produced with the interaction of a laser field, considered to be in a coherent state, with nonlinear media \cite{livroqo,livrobase,expestadocomprimido}. However, due to lack of \textit{a priori} knowledge of the laser absolute phase, the quantum state of a laser field can be written as a statistical mixture of coherent states with random phases, which is equivalent to a statistical mixture of Fock states \cite{livroqo,molmer}. So to consider that a laser field is in a coherent state is an approximation as good as to consider that it is in a Fock state. In fact, it is possible to explain experiments such as the interference produced by the superposition of two independent laser fields \cite{pfleegor67} by considering the laser fields in Fock states, since photon detections generate coherence between the two fields \cite{molmer}. This is analogous to the fact that two Bose-Einstein condensates can present interference even if both are initially in number states, since the detection of atoms generates coherence between them \cite{javanainen96,cirac96,castin97,laloe07}. The direct measurement of the electric field of light waves \cite{Goulielmakis1267} can also be explained by considering the laser initially in a Fock state \cite{artigoPablo}. In this case, the $f:2f$ interferometer used to calibrate the carrier–envelope phase of the light pulses in the experiments induces the appearance of a coherent superposition of Fock states in the system, generating a quantum state whose expectation value of the electric field operator is nonzero \cite{artigoPablo}.

Here we describe experiments that produce and characterize single-mode squeezed vacuum states of light considering this mixed state description of the laser field, not being in a coherent state. In the Fock state basis, the experiments are described without the production of squeezed states of light in any part of the setup and without the concept of squeezing being used in the analysis. By considering a Fock state for the laser, entanglement is produced between the signal mode that exits the nonlinear medium and the mode used as the local oscillator in homodyne detection. The interference of these modes at a beam splitter produce the same experimental predictions as the consideration of a coherent state for the laser, with the production of a squeezed-light state in the setup which is submitted to homodyne detection. The difference in interpretations of the experimental results when different basis (coherent or Fock states) are used to describe the laser field state is discussed.  We provide a general physical explanation for the noise reduction in the experiments in terms of a better definition of the relative phase between the signal field and the field used as local oscillator.

\begin{figure}
  \centering
    \includegraphics[width=8.5cm]{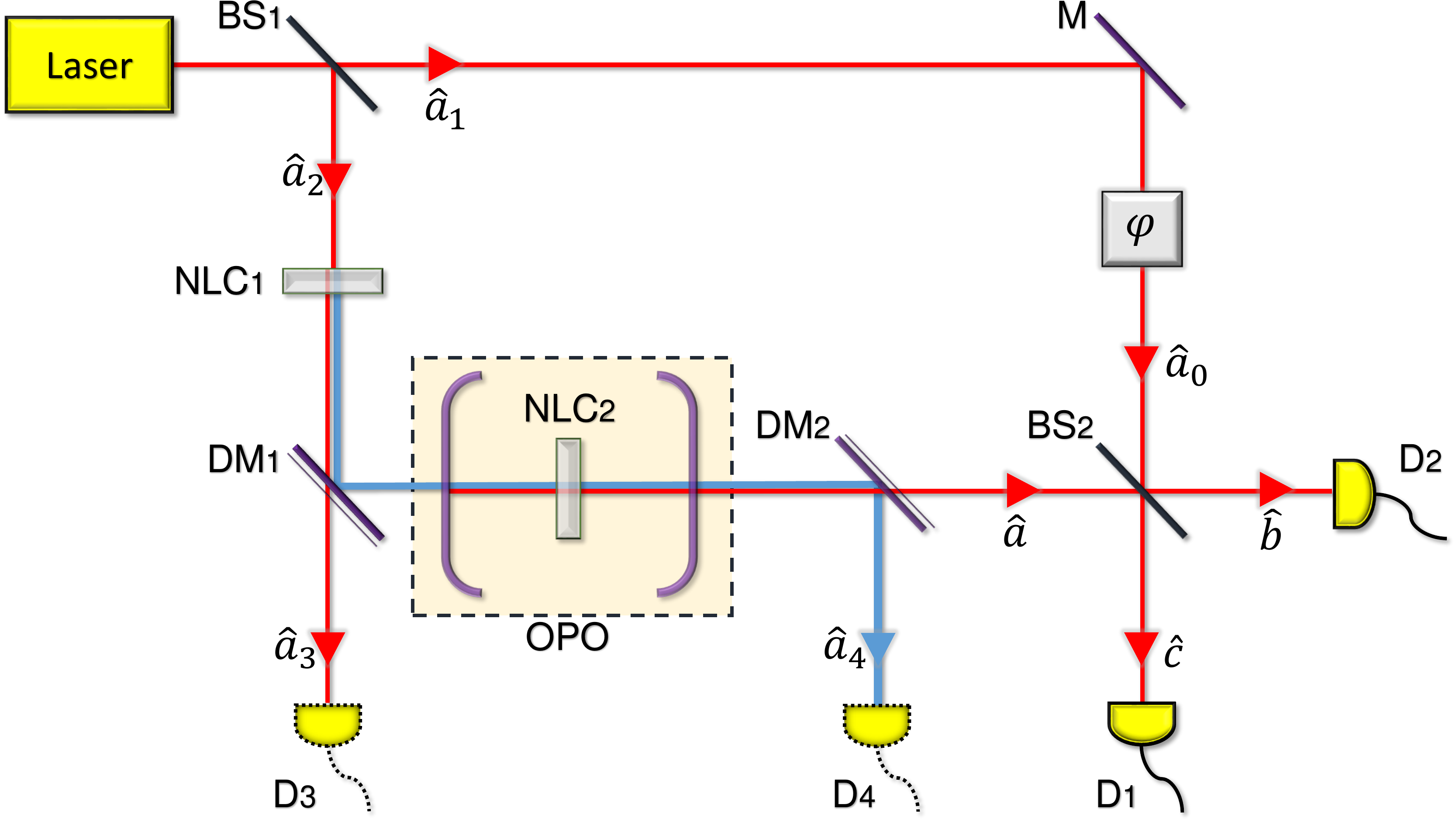}
  \caption{Scheme for producing and characterizing a single-mode squeezed vacuum state. BS(1,2) - beam splitters, NLC(1,2) - nonlinear crystals, DM(1,2) - dichroic mirrors, OPO - optical parametric oscillator, M - mirror, D(1,2,3,4) - detectors, $\varphi$ - adjustable phase.}\label{fig:interferometer}
\end{figure}

The scheme showed in Fig. \ref{fig:interferometer} illustrates an experiment for the generation and characterization of the single-mode squeezed vacuum state. Considering that there is a coherent state leaving the laser, coherent states with smaller amplitudes are transmitted and reflected by the beam splitter BS$_1$. The reflected field passes through the nonlinear crystal NLC$_1$, where second harmonic generation occurs. The field with frequency $2\omega$ is reflected by the dichroic mirror DM$_1$ and the field with frequency $\omega$ is transmitted. The reflected field goes to an optical parametric oscillator (OPO), generating a field with frequency $\omega$ in the squeezed vacuum state \cite{livroqo,livrobase}. The dichroic mirror DM$_2$  at the exit of the OPO reflects the $2\omega$ field and transmits the $\omega$ field to mode $a$, which is combined with a coherent state in mode $a_0$ at beam splitter BS$_2$ for the process of homodyne detection. Explicitly, the state of the signal field in mode $a$ can be written in the Fock basis as 
\begin{align}\label{eq:0}
    \begin{split}
        &\left|\xi\right>=\sum_{m=0}^{\infty}C_m\left|2m\right>_a, \text{ with}\\
        &C_m=\sqrt{\operatorname{sech}r}(-1)^m\frac{\sqrt{(2m)!}}{2^{m}m!}(\mathrm{e}^{i\phi}\operatorname{tanh}r)^{m},
    \end{split}
\end{align}
where the squeezing parameters $r$ and $\phi$ come from the squeeze operator, defined as $\hat{S}(\xi)=\mathrm{e}^{\frac{1}{2}(\xi^*\hat{a}^2-\xi\hat{a}^{\dagger2})}$ with $\xi\equiv r\mathrm{e}^{i\phi}$ \cite{livroqo,livrobase}. The action of the squeeze operator in the electromagnetic vacuum state $|\mathrm{vac}\rangle$ results in a squeezed vacuum state above: $\left|\xi\right>=\hat{S}(\xi)|\mathrm{vac}\rangle$.

Squeezed states are usually characterized by homodyne detection, like depicted in Fig. \ref{fig:interferometer}. The signal field in mode $a$ is combined with the local oscillator in a coherent state with large amplitude $|\beta\mathrm{e}^{i\varphi}\rangle_{a_0}$ (with real $\beta$ and $\varphi$) in mode $a_0$ at $\text{BS}_2$, assumed to be a symmetric 50:50 beam splitter. The intensity measured by detector D$_1$ is proportional to $\hat{c}^\dag\hat{c}$, with $\hat{c}=(\hat{a}_0+i\hat{a})/\sqrt{2}$ being the annihilation operator for the field in mode $c$ and similarly for the other modes, while the intensity measured by detector D$_2$ is proportional to $\hat{b}^\dag\hat{b}$, with $\hat{b}=(\hat{a}+i\hat{a}_0)/\sqrt{2}$. Defining $\hat{n}_{bc}=\hat{n}_b-\hat{n}_c$, under the above circumstances we have 
\begin{equation}\label{hom}
	\left<\hat{n}_{bc}\right>\approx 2 \beta\left<\hat{X}(\theta)\right>,\;\left<(\Delta\hat{n}_{bc})^2\right>\approx4\beta^2\left<(\Delta\hat{X}(\theta))^2\right>,
\end{equation}
where $\hat{X}(\theta)\equiv(\hat{a}\rme^{-i\theta}+\hat{a}^\dag\rme^{i\theta})/2$ is a quadrature operator for the field in mode $a$ and $\theta=\varphi+\pi/2$ \cite{livroqo,livrobase}. So, by varying the parameter $\varphi$ of the local oscillator, the expectation value and variances of different quadratures can be obtained. For the squeezed vacuum state of Eq. (\ref{eq:0}) we have \cite{livroqo,livrobase}
\begin{eqnarray}\label{var}\nonumber
	 \left<(\Delta\hat{n}_{bc})^2\right>\approx&&\beta^2[\cosh^2{r}+\sinh^2{r}-\\
    &&-2\sinh{r}\cosh{r}\cos{(2\theta-\phi)}].
\end{eqnarray}
According to Eq. (\ref{hom}), variances for $\hat{n}_{bc}$ smaller than $\beta^2$ imply variances for $\hat{X}(\theta)$ smaller than 1/4, characterizing squeezing. 

However, as mentioned before, due to lack of \textit{a priori} knowledge of the laser absolute phase, the laser field state can be written as a statistical mixture of coherent states or, similarly, by a statistical mixture of Fock states \cite{livroqo,molmer}:
\begin{align}\label{eq:2}
\rho_l=\int_0^{2\pi}\frac{d\phi'}{2\pi}\left|\alpha\mathrm{e}^{i\phi'}\right>\left<\alpha\mathrm{e}^{i\phi'}\right|=\sum_{n=0}^{\infty}P_n\left|n\right>\left<n\right|,
\end{align}
with $\alpha$ real and $P_n=\alpha^{2n}e^{-\alpha^2}/n!$ giving a Poissonian probability distribution for the number of photons with mean number $\alpha^2$ \cite{livroqo,molmer}. Note that if the laser is in a coherent state $|\alpha\mathrm{e}^{i\phi'}\rangle$, the phase $\phi'$ defines the phase $\phi$ of the squeezed state of Eq. (\ref{eq:0}). So if we have an incoherent combination of absolute phases for the laser field as in Eq. (\ref{eq:2}), the incoherent combination of squeezed quadratures results in a non-squeezed state for the field in mode $a$ of Fig. 1. The experimental quantum teleportation using squeezed states \cite{teleportation} was previously criticized due to the fact that, since the laser field is not a coherent state, the experimental results could not be associated to a quantum teleportation protocol \cite{artigocontinuousvariableteleportation2001}. After an intense debate \cite{artigoPhysRevLettQuantumStateLaser,artigoPhysRevAPhotonNumberSuperselection,artigotwoviewsonquantumcoherence,artigoReviewModern}, an interesting point of view was constructed claiming that a phase reference frame must be established for any quantum information protocol in continuous variables, and under this perspective the experimental results of Ref. \cite{teleportation} do demonstrate the implementation of a quantum teleportation protocol \cite{artigoPhysRevAPhotonNumberSuperselection,artigotwoviewsonquantumcoherence}. This discussion is related to the question about if coherence as the quantum superposition of energy eigenstates in the optical regime (as in a coherent state) is a  \textit{fiction}, as initially argued by M\o lmer \cite{molmer}, or a \textit{fact}.  But if we consider that a quantum state refers not only to the intrinsic properties of a system, but also to its relation to external parts, there is no dilemma. In this view, coherence as a fiction is associated to an internal treatment of the phase reference frame, while coherence as a fact is associated to an external treatment of the phase reference frame, both views being valid \cite{artigotwoviewsonquantumcoherence}. In the scheme of Fig. 1, note that the phase $\phi'$ of the laser field also defines the phase $\varphi$ of the local oscillator in mode $a_0$ of Fig. 1. So, with respect to the phase of the local oscillator, the quantum state of the field in mode $a$ of Fig. 1 is a squeezed state. Since the local oscillator is the responsible for establishing the phase reference, in this perspective the experiments represented by the setup of Fig. 1 do demonstrate squeezing. 

However, according to Eq. (\ref{eq:2}), to consider that a laser field is in a coherent state is an approximation as good as to consider that it is in a Fock state. In the following we show how it is possible to arrive at the same experimental predictions of Eq. (\ref{var}) by considering the laser field in a Fock state, but with a completely different description of the phenomenon.

If a Fock state $\left|M\right>$ is considered to be the output state of the laser field in the setup of Fig. 1, the state right after the beam splitter BS$_1$ is an entangled state of the form $\left|\psi_0\right>=\sum_{s=0}^{\infty}A_s\left|M-s\right>_{a_1}\left|s\right>_{a_2}$,
where $s$ is the number of reflected photons, $M-s$ is the number of transmitted photons, and $A_s$ is a complex coefficient. Subsequently, at the $\text{NLC}_1$ second harmonic generation occurs and the photons with frequency $2\omega$ go to the OPO, where parametric down-conversion happens. Since in the parametric down-conversion process photons of frequency $2\omega$ are converted into pairs of photons of frequency $\omega$, only even photon numbers are allowed in the $a$ mode. Just to simplify the discussion, we consider the presence of two perfect detectors $\text{D}_3$ and $\text{D}_4$ that count the photon numbers $N_{a_3}$ and $N_{a_4}$ that exit through the $a_3$ and $a_4$ modes. The total photon number at the modes $a$ and $a_0$ is therefore $N=M-N_{a_3}-2N_{a_4}$, in such a way that the state just before the beam splitter BS$_2$ is
\begin{align}\label{eq:4}
    \left|\psi\right>=\sum_{m=0}^{\infty}C_m\left|2m\right>_{a}\left|N-2m\right>_{a_0}\rme^{i(N-2m)\varphi},
\end{align}
where $C_m$ is the probability amplitude of generating $m$ photon pairs at the OPO. These coefficients $C_m$ are the same as the ones from Eq. (\ref{eq:0}), since the probability amplitude of generating a photon pair should not depend if the incident field at the OPO is a coherent state with an uncertainty in the number of photons much smaller than the average number of photons, as considered in the deduction of Eq. (\ref{eq:0}), or if it is a combination of Fock states with the same average and a small uncertainty in the photon number, as considered in the deduction of Eq. (\ref{eq:4}). Considering also that the number of photons in mode $a$ is much smaller than $N$, as necessary in the process of homodyne detection, in the Appendix  we arrive at the following result for the variance of the difference in the number of photons detected by D$_1$ and D$_2$: 
\begin{align}\label{eq:6}
    \begin{split}
        \left<(\Delta\hat{n}_{bc})^{2}\right>&\approx N[\cosh^2r+\sinh^2r-\\
        &-2\cosh{r}\sinh{r}\cos{(2\theta-\phi)}],        
    \end{split}
\end{align}
with $\theta=\varphi+\pi/2$. This result is equivalent to the one of Eq. \eqref{var} for $N\approx\beta^2$. 

Note that, by considering the laser field in a Fock state, by using Eq. (\ref{eq:4}) we conclude that the quantum state of the field in the $a$ mode is not a squeezed state, but a statistical mixture of even photon numbers of the form $\sum_{m=0}^{\infty}|C_m|^2|2m\rangle{\langle2m|}$. So the physical explanation of the phenomenon described in Fig. 1 in the Fock state basis does not involve squeezing. The experimental results can be associated to the entangled state between modes $a$ and $a_0$ described in Eq. (\ref{eq:4}). For each pair of photons detected by detectors D$_1$ and D$_2$, there is a fundamental indistinguishability about if they came directly from the laser to mode $a_0$ or if they were converted by the NLC$_1$ into a photon of frequency $2\omega$ and converted by the OPO into a pair of photons with frequency $\omega$ again, arriving at BS$_2$ through mode $a$. The interference between these two probability amplitudes results in the noise increase or decrease represented in Eq. (\ref{eq:6}), depending on the phase $\varphi$ of the interferometer.  

Essentially the same result of Eq. (\ref{eq:6}) is obtained when the more realistic description of the laser field from Eq. (\ref{eq:2}) is employed, which consists on a statistical mixture of Fock states. If the number of photons of the initial laser field is unknown and the number of photons that exit trough modes $a_3$ and $a_4$ is also unknown, there will be a probability distribution on the value of $N$ in Eq. (\ref{eq:6}), with an average value $\bar{N}$ and a standard deviation $\Delta N$. But since we have $\Delta N/\bar{N}\propto1/\sqrt{N}\rightarrow 0$ in this situation, the results are essentially the same with the substitution $N\rightarrow\bar{N}$ en Eq. \eqref{eq:6}. 

Let us comment now on how the physical explanations of the experimental results of the setups of Fig. 1 are directly related to the basis chosen to describe the laser quantum state from Eq. (\ref{eq:2}), and how these physical explanations are inconsistent one with the other. In the coherent state basis, a squeezed state is produced in the OPO of Fig. 1 and a homodyne measurement is performed on the field $a$ with the field $a_0$ acting as a local oscillator. Even if the laser field does not have a defined absolute phase, being in the mixed state of Eq. (\ref{eq:2}), the phase of the local oscillator acts as a reference phase. But note that this explanation is inconsistent with the description of the laser field in the Fock state basis, since in this case there can be no coherent superposition of different Fock states in the field of mode $a$, only statistical mixtures of Fock states being possible, such that there can be no squeezing. If, on the other hand, we use the Fock state basis to describe the laser field, the formation of entangled states like the one of Eq. (\ref{eq:4}) are responsible for explaining the experimental results of the setup of Fig. 1. Since the results of Eq. (\ref{eq:6}) do not have a strong dependence on the number of detected photons $N$, if the laser state has a small relative uncertainty in the number of photons, like in the mixed state of Eq.  (\ref{eq:2}), the results are essentially the same. But note that this explanation is inconsistent with the description of the laser field state in the coherent state basis, since the incidence of a coherent state on a beam splitter results in a separate state for the transmitted and reflected modes, such that no entanglement as expressed in Eq. (\ref{eq:4}) can occur. So none of these presented explanations is general, since each one works only in a particular basis for the laser field representation.

We now provide a general explanation for the phenomenon, valid in any basis, related to the variance of the relative phase distribution between the signal field and the local oscillator field. There are different arguments that lead to following expression for the relative phase distribution between two modes of the same frequency in a state $\rho$ \cite{luis96,sanders95,luis96b}:
\begin{align}\label{Phase difference probability}
    P(\Phi)=\sum_{N=0}^\infty\langle\Phi^{(N)}|\,\rho\,|\Phi^{(N)}\rangle,\,\,\,\,\mathrm{with}
\end{align} 
\begin{align}\label{eigenstates phase difference operator}
     |\Phi^{(N)}\rangle=\frac{1}{\sqrt{2\pi}}\sum_{n_1=0}^N \mathrm{e}^{in_1\Phi}|n_1\rangle_a |N-n_1\rangle_{a_0}.
\end{align}
The states $\left|\Phi^{(N)}\right>$ are common eigenstates of the total photon number (with eigenvalue $N$) and different phase difference operators (with eigenvalue $\Phi$) among the two considered modes \cite{barnett90,luis93,luis96}. They are also associated to optimum phase measurements in interferometers \cite{sanders95,luis96b}. Substituting the quantum state of Eq. (\ref{eq:4}) in Eq. (\ref{Phase difference probability}), we have
\begin{align}\label{prob.dist.Fock result}
    \begin{split}
        P_\psi(\Phi)=\frac{1}{2\pi}\Bigg|\sum_{m=0}^{m_{max}} C_m\mathrm{e}^{-2im(\Phi+\varphi)}\Bigg|^2,
    \end{split}
\end{align}
with $m_{max}\approx N/2$. If, as we are considering here, the number of photons in mode $a$ is much smaller than $N$, such that $|C_m|$ is negligible when $2m$ is of the order of $N$, we can consider $m_{max}\rightarrow\infty$, such that $P_\psi(\Phi)$ is independent of $N$ in this limit. So, when the pure state of Eq. (\ref{eq:4}) is substituted by a mixed state with small $\Delta N/N$, we have the same phase difference distribution of Eq. (\ref{prob.dist.Fock result}). Note also that, by considering the problem in the coherent state basis for the laser, Eq. (\ref{Phase difference probability}) will make a projection in an eigenstate of the total number of photons in modes $a_0$ and $a$, such that following the adopted approximations we also arrive in the phase difference distribution of Eq. (\ref{prob.dist.Fock result}). 

In Fig. 2(a) we plot the phase difference distribution of Eq. (\ref{prob.dist.Fock result}) when the coefficients $C_m$ are given by Eq. (\ref{eq:0}) with $\phi=0$, $r=1$ and $r=2$, considering also $\varphi=0$. There are two peaks in these distributions, at $\Phi=\pi/2$ and $\Phi=3\pi/2$, whose widths decrease with the increase of $r$. Actually, the variance of each peak has the same dependence with $r$ as the standard deviation of the squeezed quadrature for the squeezed vacuum state of Eq. (\ref{eq:0}), which decay as  $\mathrm{e}^{-r}$ \cite{livroqo,livrobase}. Figure 2(b) demonstrates this behavior, where is plotted the logarithm of the ratio between the variance of $P_\psi(\Phi)$, $\sigma_\psi^2$, and the variance when the mode $a$ is in the vacuum state, $\sigma_0^2$, in a function of $r$, considering only the region between $\Phi=0$ and $\Phi=\pi$. The dashed green line corresponds to a dependence $\sigma_\psi^2=\sigma_0^2\mathrm{e}^{-r}$. Since, by Eq. (\ref{hom}), the standard deviation of $\hat{n}_{bc}$ is proportional to the standard deviation of $\hat{X}(\theta)$, its minimum value is proportional to $\sigma_\psi^2$. So a general explanation for the decrease of the standard deviation of $\hat{n}_{bc}$ with an increase of the parameter $r$, valid in any basis description of the laser field state, is a decrease in the variance of the relative phase of the fields in modes $a$ and $a_0$ in Fig. \ref{fig:interferometer}.

\begin{figure}
    \centering
    \includegraphics[width=8.5cm]{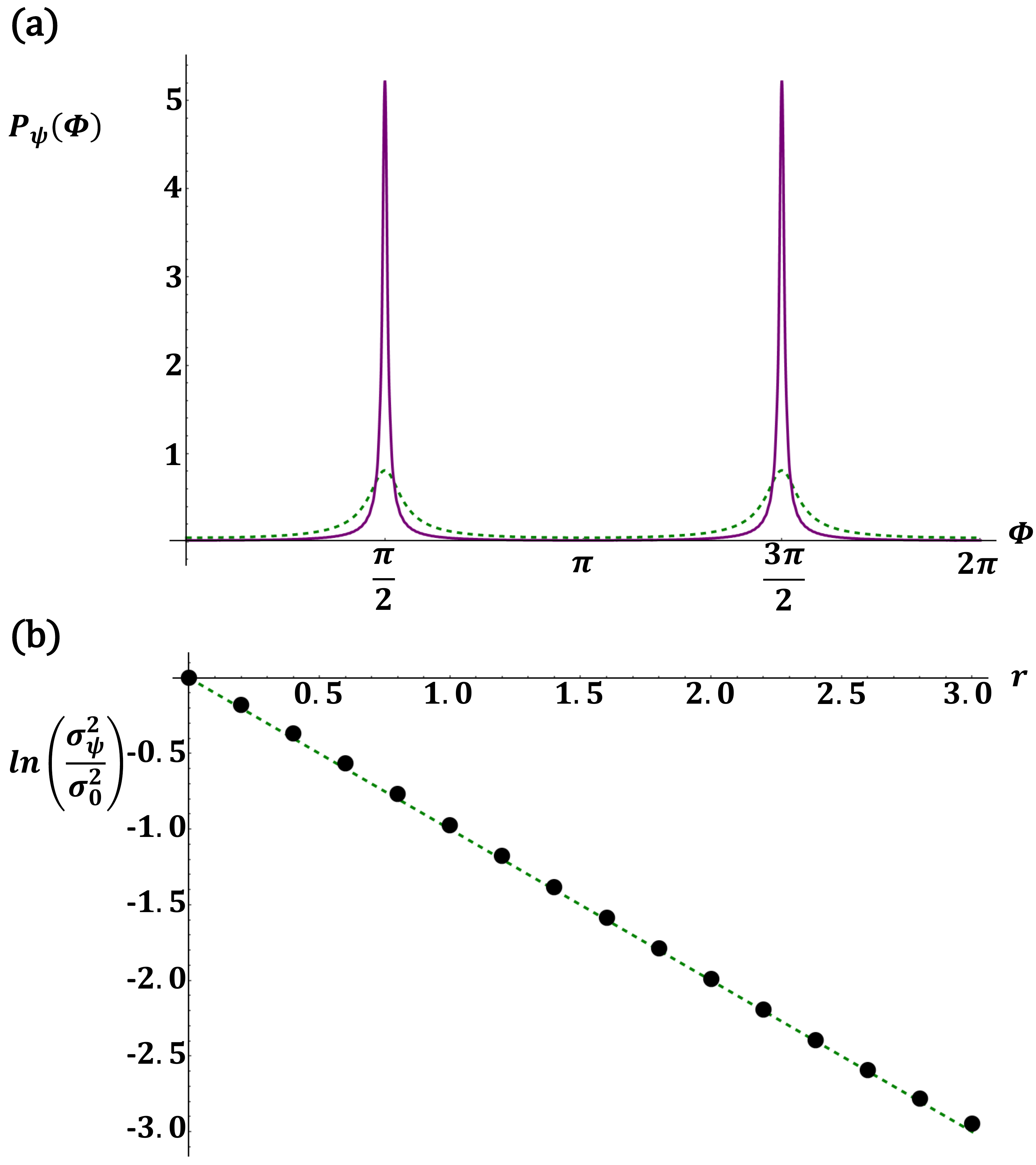}
    \caption{Probability distributions for the phase difference between signal and local oscillator fields in squeezed vacuum states. (a) $P_\psi(\Phi)$ from Eq. (\ref{prob.dist.Fock result}) for $\varphi=0$ and $C_m$  given by Eq. (\ref{eq:0}) with $\phi=0$, $r=1$ (dashed green line) and $r=2$ (solid purple line). b) Logarithm of the ratio between the variance of $P_\psi(\Phi)$ from Eq. (\ref{prob.dist.Fock result}), $\sigma_\psi^2$, and the variance when the mode $a$ is in the vacuum state, $\sigma_0^2$, in function of $r$, considering only the region between $\Phi=0$ and $\Phi=\pi$. The dashed green line corresponds to a dependence $\sigma_\psi^2=\sigma_0^2\mathrm{e}^{-r}$. We used $m_{max}=200$ in Eq. (\ref{prob.dist.Fock result}) for all plots.}
    \label{fig:DistProbFaseR1R2}
\end{figure}


To summarize, we have shown how squeezed vacuum experiments can be described without the production of squeezed-light states and without the concept of light squeezing being used in the treatment, when the laser field state is described in the Fock basis. We conclude that a general explanation for the the decrease on the fluctuations of the difference of intensities measured by detectors $D_1$ and $D_2$ in the scheme of Fig. \ref{fig:interferometer} with an increase of the squeezing parameter $r$ is a better definition of the phase difference between the fields in modes $a$ and $a_0$. This explanation is valid in any basis we use to describe the laser field state, contrary to the vacuum squeezed state generation by the OPO, which is valid only in the coherent state basis for the laser state, and the entanglement generation between the modes $a$ and $a_0$, which is valid only in the Fock state basis for the laser state. We hope that this deeper understanding we provide for the phenomenon of light squeezing, with so many applications, can inspire fruitful advances in this field. 

The authors acknowledge Paulo Nussenzveig and Marcelo Martinelli for very useful discussions. This work was supported by the Brazilian agencies CNPq and CAPES.

\appendix* \section{}

In this appendix we show that the variance of $\hat{n}_{bc}\equiv\hat{n}_b-\hat{n}_c=i(\hat{a}^\dagger\hat{a}_0+\hat{a}_0^\dagger\hat{a}$) for the state of Eq. (\ref{eq:4}) is given by Eq.  (\ref{eq:6}) when the number of photons in mode $a$ is much smaller than $N$. Using the state of Eq. \eqref{eq:4} and making the approximation $N-2m\approx N-2m-1\approx N$, one obtains
\begin{align}
    \begin{split}
        &\left<(\Delta \hat{n}_{bc})^2\right>\approx N\sum_{m=0}^\infty \Bigg\{|C_m|^2(4m+1)+\\
        &+2\mathrm{Re}\left[C^*_mC_{m+1}\sqrt{(2m+2)(2m+1)}\mathrm{e}^{-2i\theta}\right]\Bigg\},
    \end{split}
\end{align}
where $\theta=\varphi+\pi/2$. Substituting the coefficients $C_m$ from Eq. (\ref{eq:0}), we have 
\begin{equation}\label{Dnbc}
	\left<(\Delta \hat{n}_{bc})^2\right>\approx N \left[A-B\cos{(2\theta-\phi)}\right],
\end{equation}
with
\begin{eqnarray}\nonumber
        && A=\frac{1}{\mathrm{cosh}r}\sum_{m=0}^\infty\frac{(2m)!}{2^{2m}(m!)^2}(\operatorname{tanh}^2r)^{m}(1+4m),\\\nonumber
        && B=\frac{\mathrm{sinh}r}{\mathrm{cosh}^2r}\sum_{m=0}^\infty\frac{(2m)!}{2^{2m}(m!)^2}(\operatorname{tanh}^2r)^{m}(2+4m).
\end{eqnarray}
Using the relations
\begin{eqnarray}\label{expansao}\nonumber
   && (1-x)^{-1/2}=\sum_{m=0}^{\infty}\frac{(2m)!}{2^{2m}(m!)^2}x^m,\\\nonumber
		&& \frac{d}{dx}(1-x)^{-1/2}=\sum_{m=0}^{\infty}\frac{(2m)!}{2^{2m}(m!)^2}x^{m-1}m=\frac{1}{2}(1-x)^{-3/2},
\end{eqnarray}
with $x=\operatorname{tanh}^2r$, after some manipulations we arrive at values for $A$ and $B$ that lead Eq. (\ref{Dnbc}) to Eq. (\ref{eq:6}).


\begin{thebibliography}{35}%
\makeatletter
\providecommand \@ifxundefined [1]{%
 \@ifx{#1\undefined}
}%
\providecommand \@ifnum [1]{%
 \ifnum #1\expandafter \@firstoftwo
 \else \expandafter \@secondoftwo
 \fi
}%
\providecommand \@ifx [1]{%
 \ifx #1\expandafter \@firstoftwo
 \else \expandafter \@secondoftwo
 \fi
}%
\providecommand \natexlab [1]{#1}%
\providecommand \enquote  [1]{``#1''}%
\providecommand \bibnamefont  [1]{#1}%
\providecommand \bibfnamefont [1]{#1}%
\providecommand \citenamefont [1]{#1}%
\providecommand \href@noop [0]{\@secondoftwo}%
\providecommand \href [0]{\begingroup \@sanitize@url \@href}%
\providecommand \@href[1]{\@@startlink{#1}\@@href}%
\providecommand \@@href[1]{\endgroup#1\@@endlink}%
\providecommand \@sanitize@url [0]{\catcode `\\12\catcode `\$12\catcode
  `\&12\catcode `\#12\catcode `\^12\catcode `\_12\catcode `\%12\relax}%
\providecommand \@@startlink[1]{}%
\providecommand \@@endlink[0]{}%
\providecommand \url  [0]{\begingroup\@sanitize@url \@url }%
\providecommand \@url [1]{\endgroup\@href {#1}{\urlprefix }}%
\providecommand \urlprefix  [0]{URL }%
\providecommand \Eprint [0]{\href }%
\providecommand \doibase [0]{http://dx.doi.org/}%
\providecommand \selectlanguage [0]{\@gobble}%
\providecommand \bibinfo  [0]{\@secondoftwo}%
\providecommand \bibfield  [0]{\@secondoftwo}%
\providecommand \translation [1]{[#1]}%
\providecommand \BibitemOpen [0]{}%
\providecommand \bibitemStop [0]{}%
\providecommand \bibitemNoStop [0]{.\EOS\space}%
\providecommand \EOS [0]{\spacefactor3000\relax}%
\providecommand \BibitemShut  [1]{\csname bibitem#1\endcsname}%
\let\auto@bib@innerbib\@empty
\bibitem [{\citenamefont {{W. F. Walls}}\ and\ \citenamefont {{G. J.
  Milburn}}(2008)}]{livroqo}%
  \BibitemOpen
  \bibfield  {author} {\bibinfo {author} {\bibnamefont {{W. F. Walls}}}\ and\
  \bibinfo {author} {\bibnamefont {{G. J. Milburn}}},\ }\href@noop {} {\emph
  {\bibinfo {title} {Quantum Optics}}}\ (\bibinfo  {publisher} {Springer-Verlag
  Berlin Heidelberg},\ \bibinfo {year} {2008})\BibitemShut {NoStop}%
\bibitem [{\citenamefont {{C. C. Gerry}}\ and\ \citenamefont {{P. L.
  Knight}}(2008)}]{livrobase}%
  \BibitemOpen
  \bibfield  {author} {\bibinfo {author} {\bibnamefont {{C. C. Gerry}}}\ and\
  \bibinfo {author} {\bibnamefont {{P. L. Knight}}},\ }\href@noop {} {\emph
  {\bibinfo {title} {Introductory Quantum Optics}}}\ (\bibinfo  {publisher}
  {Cambridge University Press},\ \bibinfo {year} {2008})\BibitemShut {NoStop}%
\bibitem [{\citenamefont {Breitenbach}\ \emph {et~al.}(1997)\citenamefont
  {Breitenbach}, \citenamefont {Schillerand},\ and\ \citenamefont
  {Mlynek}}]{expestadocomprimido}%
  \BibitemOpen
  \bibfield  {author} {\bibinfo {author} {\bibfnamefont {G.}~\bibnamefont
  {Breitenbach}}, \bibinfo {author} {\bibfnamefont {S.}~\bibnamefont
  {Schillerand}}, \ and\ \bibinfo {author} {\bibfnamefont {J.}~\bibnamefont
  {Mlynek}},\ }\bibfield  {title} {\enquote {\bibinfo {title} {Measurement of
  the quantum states of squeezed light},}\ }\href {\doibase 10.1038/387471a0}
  {\bibfield  {journal} {\bibinfo  {journal} {Nature}\ }\textbf {\bibinfo
  {volume} {387}},\ \bibinfo {pages} {471} (\bibinfo {year}
  {1997})}\BibitemShut {NoStop}%
\bibitem [{\citenamefont {Caves}(1981)}]{caves81}%
  \BibitemOpen
  \bibfield  {author} {\bibinfo {author} {\bibfnamefont {C.~M.}\ \bibnamefont
  {Caves}},\ }\bibfield  {title} {\enquote {\bibinfo {title}
  {Quantum-mechanical radiation-pressure fluctuations in an interferometer},}\
  }\href {\doibase 10.1103/PhysRevLett.45.75} {\bibfield  {journal} {\bibinfo
  {journal} {Phys. Rev. Lett.}\ }\textbf {\bibinfo {volume} {45}},\ \bibinfo
  {pages} {75} (\bibinfo {year} {1980})}\BibitemShut {NoStop}%
\bibitem [{\citenamefont {Schnabel}(2017)}]{artigoPhysReports}%
  \BibitemOpen
  \bibfield  {author} {\bibinfo {author} {\bibfnamefont {R.}~\bibnamefont
  {Schnabel}},\ }\bibfield  {title} {\enquote {\bibinfo {title} {Squeezed
  states of light and their applications in laser interferometers},}\ }\href
  {\doibase https://doi.org/10.1016/j.physrep.2017.04.001} {\bibfield
  {journal} {\bibinfo  {journal} {Phys. Rep.}\ }\textbf {\bibinfo {volume}
  {684}},\ \bibinfo {pages} {1 } (\bibinfo {year} {2017})}\BibitemShut
  {NoStop}%
\bibitem [{\citenamefont {{J. Aasi \textit{et al.}}}(2013)}]{artigonaturepht}%
  \BibitemOpen
  \bibfield  {author} {\bibinfo {author} {\bibnamefont {{J. Aasi \textit{et
  al.}}}},\ }\bibfield  {title} {\enquote {\bibinfo {title} {Enhanced
  sensitivity of the ligo gravitational wave detector by using squeezed states
  of light},}\ }\href {\doibase 10.1038/nphoton.2013.177} {\bibfield  {journal}
  {\bibinfo  {journal} {Nat. Photonics}\ }\textbf {\bibinfo {volume} {7}},\
  \bibinfo {pages} {613} (\bibinfo {year} {2013})}\BibitemShut {NoStop}%
\bibitem [{\citenamefont {Vahlbruch}\ \emph {et~al.}(2016)\citenamefont
  {Vahlbruch}, \citenamefont {Mehmet}, \citenamefont {Danzmann},\ and\
  \citenamefont {Schnabel}}]{artigoPRLsqueezed}%
  \BibitemOpen
  \bibfield  {author} {\bibinfo {author} {\bibfnamefont {H.}~\bibnamefont
  {Vahlbruch}}, \bibinfo {author} {\bibfnamefont {M.}~\bibnamefont {Mehmet}},
  \bibinfo {author} {\bibfnamefont {K.}~\bibnamefont {Danzmann}}, \ and\
  \bibinfo {author} {\bibfnamefont {R.}~\bibnamefont {Schnabel}},\ }\bibfield
  {title} {\enquote {\bibinfo {title} {Detection of 15 db squeezed states of
  light and their application for the absolute calibration of photoelectric
  quantum efficiency},}\ }\href {\doibase 10.1103/PhysRevLett.117.110801}
  {\bibfield  {journal} {\bibinfo  {journal} {Phys. Rev. Lett.}\ }\textbf
  {\bibinfo {volume} {117}},\ \bibinfo {pages} {110801} (\bibinfo {year}
  {2016})}\BibitemShut {NoStop}%
\bibitem [{\citenamefont {{M. Tse \textit{et al.}}}(2019)}]{tse19}%
  \BibitemOpen
  \bibfield  {author} {\bibinfo {author} {\bibnamefont {{M. Tse \textit{et
  al.}}}},\ }\bibfield  {title} {\enquote {\bibinfo {title} {Quantum-enhanced
  advanced ligo detectors in the era of gravitational-wave astronomy},}\ }\href
  {\doibase 10.1103/PhysRevLett.123.231107} {\bibfield  {journal} {\bibinfo
  {journal} {Phys. Rev. Lett.}\ }\textbf {\bibinfo {volume} {123}},\ \bibinfo
  {pages} {231107} (\bibinfo {year} {2019})}\BibitemShut {NoStop}%
\bibitem [{\citenamefont {{F. Acernese \textit{et al.}}}(2019)}]{acernese19}%
  \BibitemOpen
  \bibfield  {author} {\bibinfo {author} {\bibnamefont {{F. Acernese \textit{et
  al.}}}},\ }\bibfield  {title} {\enquote {\bibinfo {title} {Increasing the
  astrophysical reach of the advanced virgo detector via the application of
  squeezed vacuum states of light},}\ }\href {\doibase
  10.1103/PhysRevLett.123.231108} {\bibfield  {journal} {\bibinfo  {journal}
  {Phys. Rev. Lett.}\ }\textbf {\bibinfo {volume} {123}},\ \bibinfo {pages}
  {231108} (\bibinfo {year} {2019})}\BibitemShut {NoStop}%
\bibitem [{\citenamefont {{Y. Zhao \textit{et al.}}}(2020)}]{zhao20}%
  \BibitemOpen
  \bibfield  {author} {\bibinfo {author} {\bibnamefont {{Y. Zhao \textit{et
  al.}}}},\ }\bibfield  {title} {\enquote {\bibinfo {title}
  {Frequency-dependent squeezed vacuum source for broadband quantum noise
  reduction in advanced gravitational-wave detectors},}\ }\href {\doibase
  10.1103/PhysRevLett.124.171101} {\bibfield  {journal} {\bibinfo  {journal}
  {Phys. Rev. Lett.}\ }\textbf {\bibinfo {volume} {124}},\ \bibinfo {pages}
  {171101} (\bibinfo {year} {2020})}\BibitemShut {NoStop}%
\bibitem [{\citenamefont {{L. McCuller \textit{et al.}}}(2020)}]{mcculler20}%
  \BibitemOpen
  \bibfield  {author} {\bibinfo {author} {\bibnamefont {{L. McCuller \textit{et
  al.}}}},\ }\bibfield  {title} {\enquote {\bibinfo {title}
  {Frequency-dependent squeezing for advanced ligo},}\ }\href {\doibase
  10.1103/PhysRevLett.124.171102} {\bibfield  {journal} {\bibinfo  {journal}
  {Phys. Rev. Lett.}\ }\textbf {\bibinfo {volume} {124}},\ \bibinfo {pages}
  {171102} (\bibinfo {year} {2020})}\BibitemShut {NoStop}%
\bibitem [{\citenamefont {Furusawa}\ \emph {et~al.}(1998)\citenamefont
  {Furusawa}, \citenamefont {S{\o}rensen}, \citenamefont {Braunstein},
  \citenamefont {Fuchs}, \citenamefont {Kimble},\ and\ \citenamefont
  {Polzik}}]{teleportation}%
  \BibitemOpen
  \bibfield  {author} {\bibinfo {author} {\bibfnamefont {A.}~\bibnamefont
  {Furusawa}}, \bibinfo {author} {\bibfnamefont {J.~L.}\ \bibnamefont
  {S{\o}rensen}}, \bibinfo {author} {\bibfnamefont {S.~L.}\ \bibnamefont
  {Braunstein}}, \bibinfo {author} {\bibfnamefont {C.~A.}\ \bibnamefont
  {Fuchs}}, \bibinfo {author} {\bibfnamefont {H.~J.}\ \bibnamefont {Kimble}}, \
  and\ \bibinfo {author} {\bibfnamefont {E.~S.}\ \bibnamefont {Polzik}},\
  }\bibfield  {title} {\enquote {\bibinfo {title} {Unconditional quantum
  teleportation},}\ }\href {\doibase 10.1126/science.282.5389.706} {\bibfield
  {journal} {\bibinfo  {journal} {Science}\ }\textbf {\bibinfo {volume}
  {282}},\ \bibinfo {pages} {706} (\bibinfo {year} {1998})}\BibitemShut
  {NoStop}%
\bibitem [{\citenamefont {Coelho}\ \emph {et~al.}(2009)\citenamefont {Coelho},
  \citenamefont {Barbosa}, \citenamefont {Cassemiro}, \citenamefont {Villar},
  \citenamefont {Martinelli},\ and\ \citenamefont {Nussenzveig}}]{Coelho823}%
  \BibitemOpen
  \bibfield  {author} {\bibinfo {author} {\bibfnamefont {A.~S.}\ \bibnamefont
  {Coelho}}, \bibinfo {author} {\bibfnamefont {F.~A.~S.}\ \bibnamefont
  {Barbosa}}, \bibinfo {author} {\bibfnamefont {K.~N.}\ \bibnamefont
  {Cassemiro}}, \bibinfo {author} {\bibfnamefont {A.~S.}\ \bibnamefont
  {Villar}}, \bibinfo {author} {\bibfnamefont {M.}~\bibnamefont {Martinelli}},
  \ and\ \bibinfo {author} {\bibfnamefont {P.}~\bibnamefont {Nussenzveig}},\
  }\bibfield  {title} {\enquote {\bibinfo {title} {Three-color entanglement},}\
  }\href {\doibase 10.1126/science.1178683} {\bibfield  {journal} {\bibinfo
  {journal} {Science}\ }\textbf {\bibinfo {volume} {326}},\ \bibinfo {pages}
  {823} (\bibinfo {year} {2009})}\BibitemShut {NoStop}%
\bibitem [{\citenamefont {Chen}\ \emph {et~al.}(2014)\citenamefont {Chen},
  \citenamefont {Menicucci},\ and\ \citenamefont {Pfister}}]{chen14}%
  \BibitemOpen
  \bibfield  {author} {\bibinfo {author} {\bibfnamefont {M.}~\bibnamefont
  {Chen}}, \bibinfo {author} {\bibfnamefont {N.~C.}\ \bibnamefont {Menicucci}},
  \ and\ \bibinfo {author} {\bibfnamefont {O.}~\bibnamefont {Pfister}},\
  }\bibfield  {title} {\enquote {\bibinfo {title} {Experimental realization of
  multipartite entanglement of 60 modes of a quantum optical frequency comb},}\
  }\href {\doibase 10.1103/PhysRevLett.112.120505} {\bibfield  {journal}
  {\bibinfo  {journal} {Phys. Rev. Lett.}\ }\textbf {\bibinfo {volume} {112}},\
  \bibinfo {pages} {120505} (\bibinfo {year} {2014})}\BibitemShut {NoStop}%
\bibitem [{\citenamefont {Dutt}\ \emph {et~al.}(2015)\citenamefont {Dutt},
  \citenamefont {Luke}, \citenamefont {Manipatruni}, \citenamefont {Gaeta},
  \citenamefont {Nussenzveig},\ and\ \citenamefont
  {Lipson}}]{PhysRevApplied.3.044005}%
  \BibitemOpen
  \bibfield  {author} {\bibinfo {author} {\bibfnamefont {A.}~\bibnamefont
  {Dutt}}, \bibinfo {author} {\bibfnamefont {K.}~\bibnamefont {Luke}}, \bibinfo
  {author} {\bibfnamefont {S.}~\bibnamefont {Manipatruni}}, \bibinfo {author}
  {\bibfnamefont {A.~L.}\ \bibnamefont {Gaeta}}, \bibinfo {author}
  {\bibfnamefont {P.}~\bibnamefont {Nussenzveig}}, \ and\ \bibinfo {author}
  {\bibfnamefont {M.}~\bibnamefont {Lipson}},\ }\bibfield  {title} {\enquote
  {\bibinfo {title} {On-chip optical squeezing},}\ }\href {\doibase
  10.1103/PhysRevApplied.3.044005} {\bibfield  {journal} {\bibinfo  {journal}
  {Phys. Rev. Appl.}\ }\textbf {\bibinfo {volume} {3}},\ \bibinfo {pages}
  {044005} (\bibinfo {year} {2015})}\BibitemShut {NoStop}%
\bibitem [{\citenamefont {{L. F. Mu\~noz-Mart\'{\i}nez \textit{et
  al.}}}(2018)}]{PhysRevA.98.023823}%
  \BibitemOpen
  \bibfield  {author} {\bibinfo {author} {\bibnamefont {{L. F.
  Mu\~noz-Mart\'{\i}nez \textit{et al.}}}},\ }\bibfield  {title} {\enquote
  {\bibinfo {title} {Exploring six modes of an optical parametric
  oscillator},}\ }\href {\doibase 10.1103/PhysRevA.98.023823} {\bibfield
  {journal} {\bibinfo  {journal} {Phys. Rev. A}\ }\textbf {\bibinfo {volume}
  {98}},\ \bibinfo {pages} {023823} (\bibinfo {year} {2018})}\BibitemShut
  {NoStop}%
\bibitem [{\citenamefont {Israel}\ \emph {et~al.}(2019)\citenamefont {Israel},
  \citenamefont {Cohen}, \citenamefont {Song}, \citenamefont {Joo},
  \citenamefont {Eisenberg},\ and\ \citenamefont {Silberberg}}]{Israel:19}%
  \BibitemOpen
  \bibfield  {author} {\bibinfo {author} {\bibfnamefont {Y.}~\bibnamefont
  {Israel}}, \bibinfo {author} {\bibfnamefont {L.}~\bibnamefont {Cohen}},
  \bibinfo {author} {\bibfnamefont {X.~B.}\ \bibnamefont {Song}}, \bibinfo
  {author} {\bibfnamefont {J.}~\bibnamefont {Joo}}, \bibinfo {author}
  {\bibfnamefont {H.~S.}\ \bibnamefont {Eisenberg}}, \ and\ \bibinfo {author}
  {\bibfnamefont {Y.}~\bibnamefont {Silberberg}},\ }\bibfield  {title}
  {\enquote {\bibinfo {title} {Entangled coherent states created by mixing
  squeezed vacuum and coherent light},}\ }\href {\doibase
  10.1364/OPTICA.6.000753} {\bibfield  {journal} {\bibinfo  {journal} {Optica}\
  }\textbf {\bibinfo {volume} {6}},\ \bibinfo {pages} {753} (\bibinfo
  {year} {2019})}\BibitemShut {NoStop}%
\bibitem [{\citenamefont {M\o{}lmer}(1997)}]{molmer}%
  \BibitemOpen
  \bibfield  {author} {\bibinfo {author} {\bibfnamefont {K.}~\bibnamefont
  {M\o{}lmer}},\ }\bibfield  {title} {\enquote {\bibinfo {title} {Optical
  coherence: A convenient fiction},}\ }\href {\doibase
  10.1103/PhysRevA.55.3195} {\bibfield  {journal} {\bibinfo  {journal} {Phys.
  Rev. A}\ }\textbf {\bibinfo {volume} {55}},\ \bibinfo {pages} {3195}
  (\bibinfo {year} {1997})}\BibitemShut {NoStop}%
\bibitem [{\citenamefont {Pfleegor}\ and\ \citenamefont
  {Mandel}(1967)}]{pfleegor67}%
  \BibitemOpen
  \bibfield  {author} {\bibinfo {author} {\bibfnamefont {R.~L.}\ \bibnamefont
  {Pfleegor}}\ and\ \bibinfo {author} {\bibfnamefont {L.}~\bibnamefont
  {Mandel}},\ }\bibfield  {title} {\enquote {\bibinfo {title} {Interference of
  independent photon beams},}\ }\href {\doibase 10.1103/PhysRev.159.1084}
  {\bibfield  {journal} {\bibinfo  {journal} {Phys. Rev.}\ }\textbf {\bibinfo
  {volume} {159}},\ \bibinfo {pages} {1084} (\bibinfo {year}
  {1967})}\BibitemShut {NoStop}%
\bibitem [{\citenamefont {Javanainen}\ and\ \citenamefont
  {Yoo}(1996)}]{javanainen96}%
  \BibitemOpen
  \bibfield  {author} {\bibinfo {author} {\bibfnamefont {J.}~\bibnamefont
  {Javanainen}}\ and\ \bibinfo {author} {\bibfnamefont {S.~M.}\ \bibnamefont
  {Yoo}},\ }\bibfield  {title} {\enquote {\bibinfo {title} {Quantum phase of a
  bose-einstein condensate with an arbitrary number of atoms},}\ }\href
  {\doibase 10.1103/PhysRevLett.76.161} {\bibfield  {journal} {\bibinfo
  {journal} {Phys. Rev. Lett.}\ }\textbf {\bibinfo {volume} {76}},\ \bibinfo
  {pages} {161} (\bibinfo {year} {1996})}\BibitemShut {NoStop}%
\bibitem [{\citenamefont {Cirac}\ \emph {et~al.}(1996)\citenamefont {Cirac},
  \citenamefont {Gardiner}, \citenamefont {Naraschewski},\ and\ \citenamefont
  {Zoller}}]{cirac96}%
  \BibitemOpen
  \bibfield  {author} {\bibinfo {author} {\bibfnamefont {J.~I.}\ \bibnamefont
  {Cirac}}, \bibinfo {author} {\bibfnamefont {C.~W.}\ \bibnamefont {Gardiner}},
  \bibinfo {author} {\bibfnamefont {M.}~\bibnamefont {Naraschewski}}, \ and\
  \bibinfo {author} {\bibfnamefont {P.}~\bibnamefont {Zoller}},\ }\bibfield
  {title} {\enquote {\bibinfo {title} {Continuous observation of interference
  fringes from bose condensates},}\ }\href {\doibase 10.1103/PhysRevA.54.R3714}
  {\bibfield  {journal} {\bibinfo  {journal} {Phys. Rev. A}\ }\textbf {\bibinfo
  {volume} {54}},\ \bibinfo {pages} {R3714} (\bibinfo {year}
  {1996})}\BibitemShut {NoStop}%
\bibitem [{\citenamefont {Castin}\ and\ \citenamefont
  {Dalibard}(1997)}]{castin97}%
  \BibitemOpen
  \bibfield  {author} {\bibinfo {author} {\bibfnamefont {Y.}~\bibnamefont
  {Castin}}\ and\ \bibinfo {author} {\bibfnamefont {J.}~\bibnamefont
  {Dalibard}},\ }\bibfield  {title} {\enquote {\bibinfo {title} {Relative phase
  of two bose-einstein condensates},}\ }\href {\doibase
  10.1103/PhysRevA.55.4330} {\bibfield  {journal} {\bibinfo  {journal} {Phys.
  Rev. A}\ }\textbf {\bibinfo {volume} {55}},\ \bibinfo {pages} {4330}
  (\bibinfo {year} {1997})}\BibitemShut {NoStop}%
\bibitem [{\citenamefont {Lalo\"e}\ and\ \citenamefont
  {Mullin}(2007)}]{laloe07}%
  \BibitemOpen
  \bibfield  {author} {\bibinfo {author} {\bibfnamefont {F.}~\bibnamefont
  {Lalo\"e}}\ and\ \bibinfo {author} {\bibfnamefont {W.~J.}\ \bibnamefont
  {Mullin}},\ }\bibfield  {title} {\enquote {\bibinfo {title} {Nonlocal quantum
  effects with bose-einstein condensates},}\ }\href {\doibase
  10.1103/PhysRevLett.99.150401} {\bibfield  {journal} {\bibinfo  {journal}
  {Phys. Rev. Lett.}\ }\textbf {\bibinfo {volume} {99}},\ \bibinfo {pages}
  {150401} (\bibinfo {year} {2007})}\BibitemShut {NoStop}%
\bibitem [{\citenamefont {{E. Goulielmakis \textit{et
  al.}}}(2004)}]{Goulielmakis1267}%
  \BibitemOpen
  \bibfield  {author} {\bibinfo {author} {\bibnamefont {{E. Goulielmakis
  \textit{et al.}}}},\ }\bibfield  {title} {\enquote {\bibinfo {title} {Direct
  measurement of light waves},}\ }\href {\doibase 10.1126/science.1100866}
  {\bibfield  {journal} {\bibinfo  {journal} {Science}\ }\textbf {\bibinfo
  {volume} {305}},\ \bibinfo {pages} {1267} (\bibinfo {year}
  {2004})}\BibitemShut {NoStop}%
\bibitem [{\citenamefont {Saldanha}(2014)}]{artigoPablo}%
  \BibitemOpen
  \bibfield  {author} {\bibinfo {author} {\bibfnamefont {P.~L.}\ \bibnamefont
  {Saldanha}},\ }\bibfield  {title} {\enquote {\bibinfo {title} {Quantum
  analysis of the direct measurement of light waves},}\ }\href {\doibase
  10.1088/1367-2630/16/1/013021} {\bibfield  {journal} {\bibinfo  {journal}
  {New J. Phys.}\ }\textbf {\bibinfo {volume} {16}},\ \bibinfo {pages} {013021}
  (\bibinfo {year} {2014})}\BibitemShut {NoStop}%
\bibitem [{\citenamefont {Rudolph}\ and\ \citenamefont
  {Sanders}(2001)}]{artigocontinuousvariableteleportation2001}%
  \BibitemOpen
  \bibfield  {author} {\bibinfo {author} {\bibfnamefont {T.}~\bibnamefont
  {Rudolph}}\ and\ \bibinfo {author} {\bibfnamefont {B.~C.}\ \bibnamefont
  {Sanders}},\ }\bibfield  {title} {\enquote {\bibinfo {title} {Requirement of
  optical coherence for continuous-variable quantum teleportation},}\ }\href
  {\doibase 10.1103/PhysRevLett.87.077903} {\bibfield  {journal} {\bibinfo
  {journal} {Phys. Rev. Lett.}\ }\textbf {\bibinfo {volume} {87}},\ \bibinfo
  {pages} {077903} (\bibinfo {year} {2001})}\BibitemShut {NoStop}%
\bibitem [{\citenamefont {van Enk}\ and\ \citenamefont
  {Fuchs}(2001)}]{artigoPhysRevLettQuantumStateLaser}%
  \BibitemOpen
  \bibfield  {author} {\bibinfo {author} {\bibfnamefont {S.~J.}\ \bibnamefont
  {van Enk}}\ and\ \bibinfo {author} {\bibfnamefont {C.~A.}\ \bibnamefont
  {Fuchs}},\ }\bibfield  {title} {\enquote {\bibinfo {title} {Quantum state of
  an ideal propagating laser field},}\ }\href {\doibase
  10.1103/PhysRevLett.88.027902} {\bibfield  {journal} {\bibinfo  {journal}
  {Phys. Rev. Lett.}\ }\textbf {\bibinfo {volume} {88}},\ \bibinfo {pages}
  {027902} (\bibinfo {year} {2001})}\BibitemShut {NoStop}%
\bibitem [{\citenamefont {Sanders}\ \emph {et~al.}(2003)\citenamefont
  {Sanders}, \citenamefont {Bartlett}, \citenamefont {Rudolph},\ and\
  \citenamefont {Knight}}]{artigoPhysRevAPhotonNumberSuperselection}%
  \BibitemOpen
  \bibfield  {author} {\bibinfo {author} {\bibfnamefont {B.~C.}\ \bibnamefont
  {Sanders}}, \bibinfo {author} {\bibfnamefont {S.~D.}\ \bibnamefont
  {Bartlett}}, \bibinfo {author} {\bibfnamefont {T.}~\bibnamefont {Rudolph}}, \
  and\ \bibinfo {author} {\bibfnamefont {P.~L.}\ \bibnamefont {Knight}},\
  }\bibfield  {title} {\enquote {\bibinfo {title} {Photon-number superselection
  and the entangled coherent-state representation},}\ }\href {\doibase
  10.1103/PhysRevA.68.042329} {\bibfield  {journal} {\bibinfo  {journal} {Phys.
  Rev. A}\ }\textbf {\bibinfo {volume} {68}},\ \bibinfo {pages} {042329}
  (\bibinfo {year} {2003})}\BibitemShut {NoStop}%
\bibitem [{\citenamefont {Bartlett}\ \emph {et~al.}(2006)\citenamefont
  {Bartlett}, \citenamefont {Rudolph},\ and\ \citenamefont
  {Spekkens}}]{artigotwoviewsonquantumcoherence}%
  \BibitemOpen
  \bibfield  {author} {\bibinfo {author} {\bibfnamefont {S.~D.}\ \bibnamefont
  {Bartlett}}, \bibinfo {author} {\bibfnamefont {T.}~\bibnamefont {Rudolph}}, \
  and\ \bibinfo {author} {\bibfnamefont {R.~W.}\ \bibnamefont {Spekkens}},\
  }\bibfield  {title} {\enquote {\bibinfo {title} {Dialogue concerning two
  views on quantum coherence: Factist and fictionist},}\ }\href {\doibase
  10.1142/S0219749906001591} {\bibfield  {journal} {\bibinfo  {journal} {Int.
  J. Quant. Info.}\ }\textbf {\bibinfo {volume} {4}},\ \bibinfo {pages}
  {17} (\bibinfo {year} {2006})}\BibitemShut {NoStop}%
\bibitem [{\citenamefont {Bartlett}\ \emph {et~al.}(2007)\citenamefont
  {Bartlett}, \citenamefont {Rudolph},\ and\ \citenamefont
  {Spekkens}}]{artigoReviewModern}%
  \BibitemOpen
  \bibfield  {author} {\bibinfo {author} {\bibfnamefont {S.~D.}\ \bibnamefont
  {Bartlett}}, \bibinfo {author} {\bibfnamefont {T.}~\bibnamefont {Rudolph}}, \
  and\ \bibinfo {author} {\bibfnamefont {R.~W.}\ \bibnamefont {Spekkens}},\
  }\bibfield  {title} {\enquote {\bibinfo {title} {Reference frames,
  superselection rules, and quantum information},}\ }\href {\doibase
  10.1103/RevModPhys.79.555} {\bibfield  {journal} {\bibinfo  {journal} {Rev.
  Mod. Phys.}\ }\textbf {\bibinfo {volume} {79}},\ \bibinfo {pages} {555}
  (\bibinfo {year} {2007})}\BibitemShut {NoStop}%
\bibitem [{\citenamefont {Luis}\ and\ \citenamefont
  {S\'anchez-Soto}(1996)}]{luis96}%
  \BibitemOpen
  \bibfield  {author} {\bibinfo {author} {\bibfnamefont {A.}~\bibnamefont
  {Luis}}\ and\ \bibinfo {author} {\bibfnamefont {L.~L.}\ \bibnamefont
  {S\'anchez-Soto}},\ }\bibfield  {title} {\enquote {\bibinfo {title}
  {Probability distributions for the phase difference},}\ }\href {\doibase
  10.1103/PhysRevA.53.495} {\bibfield  {journal} {\bibinfo  {journal} {Phys.
  Rev. A}\ }\textbf {\bibinfo {volume} {53}},\ \bibinfo {pages} {495}
  (\bibinfo {year} {1996})}\BibitemShut {NoStop}%
\bibitem [{\citenamefont {Sanders}\ and\ \citenamefont
  {Milburn}(1995)}]{sanders95}%
  \BibitemOpen
  \bibfield  {author} {\bibinfo {author} {\bibfnamefont {B.~C.}\ \bibnamefont
  {Sanders}}\ and\ \bibinfo {author} {\bibfnamefont {G.~J.}\ \bibnamefont
  {Milburn}},\ }\bibfield  {title} {\enquote {\bibinfo {title} {Optimal quantum
  measurements for phase estimation},}\ }\href {\doibase
  10.1103/PhysRevLett.75.2944} {\bibfield  {journal} {\bibinfo  {journal}
  {Phys. Rev. Lett.}\ }\textbf {\bibinfo {volume} {75}},\ \bibinfo {pages}
  {2944} (\bibinfo {year} {1995})}\BibitemShut {NoStop}%
\bibitem [{\citenamefont {Luis}\ and\ \citenamefont {Pe\ifmmode~\check{r}\else
  \v{r}\fi{}ina}(1996)}]{luis96b}%
  \BibitemOpen
  \bibfield  {author} {\bibinfo {author} {\bibfnamefont {A.}~\bibnamefont
  {Luis}}\ and\ \bibinfo {author} {\bibfnamefont {J.}~\bibnamefont
  {Pe\ifmmode~\check{r}\else \v{r}\fi{}ina}},\ }\bibfield  {title} {\enquote
  {\bibinfo {title} {Optimum phase-shift estimation and the quantum description
  of the phase difference},}\ }\href {\doibase 10.1103/PhysRevA.54.4564}
  {\bibfield  {journal} {\bibinfo  {journal} {Phys. Rev. A}\ }\textbf {\bibinfo
  {volume} {54}},\ \bibinfo {pages} {4564} (\bibinfo {year}
  {1996})}\BibitemShut {NoStop}%
\bibitem [{\citenamefont {Barnett}\ and\ \citenamefont
  {Pegg}(1990)}]{barnett90}%
  \BibitemOpen
  \bibfield  {author} {\bibinfo {author} {\bibfnamefont {S.~M.}\ \bibnamefont
  {Barnett}}\ and\ \bibinfo {author} {\bibfnamefont {D.~T.}\ \bibnamefont
  {Pegg}},\ }\bibfield  {title} {\enquote {\bibinfo {title} {Quantum theory of
  optical phase correlations},}\ }\href {\doibase 10.1103/PhysRevA.42.6713}
  {\bibfield  {journal} {\bibinfo  {journal} {Phys. Rev. A}\ }\textbf {\bibinfo
  {volume} {42}},\ \bibinfo {pages} {6713} (\bibinfo {year}
  {1990})}\BibitemShut {NoStop}%
\bibitem [{\citenamefont {Luis}\ and\ \citenamefont
  {S\'anchez-Soto}(1993)}]{luis93}%
  \BibitemOpen
  \bibfield  {author} {\bibinfo {author} {\bibfnamefont {A.}~\bibnamefont
  {Luis}}\ and\ \bibinfo {author} {\bibfnamefont {L.~L.}\ \bibnamefont
  {S\'anchez-Soto}},\ }\bibfield  {title} {\enquote {\bibinfo {title}
  {Phase-difference operator},}\ }\href {\doibase 10.1103/PhysRevA.48.4702}
  {\bibfield  {journal} {\bibinfo  {journal} {Phys. Rev. A}\ }\textbf {\bibinfo
  {volume} {48}},\ \bibinfo {pages} {4702} (\bibinfo {year}
  {1993})}\BibitemShut {NoStop}%
\end{thebibliography}

%

\end{document}